\documentstyle[preprint,aps,axodraw]{revtex} 
\begin{document}
\draft 
\preprint{${\scriptstyle{\rm ZU-TH\ 30/97} \atop
\scriptstyle{\rm UMHEP-447}}$}
\tightenlines
\title{The DMO Sum Rule Revisited}
\author{Eugene Golowich} 
\address{Department of Physics and Astronomy,
University of Massachusetts \\
Amherst MA 01003 USA}
\author{Joachim Kambor}
\address{Institut f\"ur Theoretische Physik,
Universit\"at Z\"urich \\
CH-8057 Z\"urich, Switzerland}
\maketitle
\def\Tr{\,{\rm Tr}\,}
\def\beq{\begin{equation}}
\def\eeq{\end{equation}}
\def\beqa{\begin{eqnarray}}
\def\eeqa{\end{eqnarray}}
\begin{abstract}
\noindent
We reconsider the DMO sum rule in light of our recent two-loop 
calculations of isospin and hypercharge vector and axialvector 
current propagators in chiral perturbation theory.  A modified 
derivation valid to second order in the light quark masses is 
presented, and a phenomenological analysis yields determination 
of a combination of finite counterterms occurring in the $p^4$ and 
$p^6$ chiral lagrangians.  Suggestions are given for further study.
\end{abstract}

\vspace{1.0in}

Working in the chiral limit of massless quarks, 
Weinberg was the first to derive spectral function sum rules 
involving the vector and axialvector currents.~\cite{sw1} 
There soon followed the announcement of two additional chiral 
sum rules, one for the pion electromagnetic mass
difference~\cite{dgmly} and one (the DMO sum
rule~\cite{dmo}) involving an inverse moment of chiral 
spectral functions, both also 
derived in the chiral limit. A subsequent study of chiral 
sum rules in QCD determined that only the first Weinberg sum rule 
and the DMO sum rule survive the inclusion of quark mass 
effects.~\cite{fnr,nr}  About a decade later, the DMO sum rule was 
analyzed by applying the Gasser-Leutwyler analysis 
of chiral perturbation theory (ChPT) at one-loop 
order.~\cite{gl1,gl2,dh} 
In this paper, we report on a re-analysis of the DMO sum 
rule based on our recent calculations of vector and axialvector 
current propagators at two-loop order in ChPT.~\cite{gk1,gk4}   

\begin{center}
{\bf Derivation of DMO Sum Rule in Two-Loop ChPT}
\end{center}

Given the normalization of vector and axialvector octet 
flavour currents, 
\begin{equation}
J^\mu_{{\rm V}i} = {\bar q}{\lambda_i \over 2}\gamma^\mu q \quad 
{\rm and} \quad 
J^\mu_{{\rm A}i} = {\bar q}{\lambda_i \over 2}\gamma^\mu \gamma_5 q  
\qquad (i = 1,\ldots,8) \ \ ,
\label{dmo0}
\end{equation}
one defines the isospin vector and axialvector current propagators
respectively as 
\begin{equation}
\Delta_{{\rm k}3}^{\mu\nu}(q^2) \equiv i \int d^4x~ e^{iq\cdot x}~
\langle 0|T\left( J^\mu_{{\rm k}3} (x) J^\nu_{{\rm k}3} 
(0)\right)|0\rangle \qquad ({\rm k}= {\rm V,A}) \ .
\label{dmo1}
\end{equation}
These have the tensorial decompositions
\begin{equation}
\Delta_{{\rm k}3}^{\mu\nu}(q^2) = 
(q^\mu q^\nu - q^2 g^{\mu\nu}) \Pi_{{\rm k}3}^{(1)} (q^2 ) 
+ q^\mu q^\nu \Pi_{{\rm k}3}^{(0)} (q^2) \qquad ({\rm k} = {\rm V,A}) \ \ ,
\label{dmo2}
\end{equation}
and spectral functions obtained via imaginary parts, 
\begin{equation}
{1\over \pi} {\cal I}m~\Delta_{{\rm k}3}^{\mu\nu}(q^2) = 
(q^\mu q^\nu - q^2 g^{\mu\nu}) \rho_{{\rm k}3}^{(1)}(q^2)   
+ q^\mu q^\nu \rho_{{\rm k}3}^{(0)} (q^2) \qquad 
({\rm k} = {\rm V,A}) \ \ .
\label{dmo3}
\end{equation}
In the following we shall assume isospin symmetry, for 
which the spin-zero vector current contributions vanish 
($\Pi_{{\rm V}3}^{(0)} = 0$ and $\rho_{{\rm V}3}^{(0)} = 0$).  

The asymptotic behavior ($s \to \infty$) of the 
vector and axialvector spectral functions predicted by 
QCD is given as~\cite{fnr},\cite{nr},\cite{bnp}  
\begin{equation}
(\rho_{{\rm V}3}^{(1)} - \rho_{{\rm A}3}^{(1)} 
- \rho_{{\rm A}3}^{(0)})(s) \ \sim \ o(s^{-1})\ \ .
\label{dmo4}
\end{equation}
The information in Eq.~(\ref{dmo4}) can be used, together with 
analyticity and the corresponding asymptotic behavior of the 
polarization functions, to derive dispersion relations for the 
vector and axialvector polarization functions,
\begin{equation}
\left(\Pi_{{\rm V}3}^{(1)} - \Pi_{{\rm A}3}^{(1)}
- \Pi_{{\rm A}3}^{(0)}\right)(q^2) = \int_0^\infty ds ~ 
{(\rho_{{\rm V}3}^{(1)} - \rho_{{\rm A}3}^{(1)} 
- \rho_{{\rm A}3}^{(0)})(s) \over s - q^2 - i \epsilon} 
\ \ . \label{dmo5} 
\end{equation}

Upon evaluation at $q^2 = 0$, Eq.~(\ref{dmo2}) yields the 
inverse-moment DMO (or `$L_{10}$') sum rule~\cite{dmo},
\begin{equation}
\left(\Pi_{{\rm V}3}^{(1)} - \Pi_{{\rm A}3}^{(1)}
- \Pi_{{\rm A}3}^{(0)}\right)(0) = \int_0^\infty ds ~ 
{(\rho_{{\rm V}3}^{(1)} - \rho_{{\rm A}3}^{(1)} 
- \rho_{{\rm A}3}^{(0)})(s) \over s}\ \ .
\label{dmo6}
\end{equation}
For the purpose of phenomenological analysis, one wishes 
the spectral integral to contain only quantities which 
have already been measured.  Therefore, we make use of 
the ChPT analysis of Ref.~\cite{gkPRL} to cast the DMO 
sum rule in the optimal form, 
\begin{equation}
\left(\Pi_{{\rm V}3}^{(1)} - {\hat \Pi}_{{\rm A}3}^{(1)}
+ {d{\hat \Pi}_{{\rm A}3}^{(0)} \over dq^2} \right)(0) 
+ 4 H_{1}^{(0)}(\mu) - 2 L_{10}^{(0)}(\mu) 
= \int_0^\infty ds ~ 
{(\rho_{{\rm V}3}^{(1)} - \rho_{{\rm A}3}^{(1)})(s) \over s}\ \ ,
\label{dmo7}
\end{equation}
where $H_{1}^{(0)}(\mu)$ and $L_{10}^{(0)}(\mu)$ are 
counterterms from the $p^4$ chiral lagrangian 
and ${\hat \Pi}_{{\rm A}3}^{(1,0)}$ are finite polarization 
functions entering at two-loop order~\cite{gk4,gkPRL}.  

\begin{center}
{\bf ChPT Contributions to DMO Sum Rule}
\end{center}

From Refs.~{\cite{gk1,gk4}}, we obtain the following 
expressions for terms appearing on the left-hand-side (LHS) of 
Eq.~(\ref{dmo7}),
\begin{equation}
{\rm LHS} \ = \ G^{(1)} + G^{(2)} \ \equiv \ G^{(1)} + 
G_\pi^{(2)} + G_K^{(2)} + G_{\rm NUM}^{(2)} + G_{\rm CT}^{(2)} \ \ ,
\label{dmo8}
\end{equation}
with 
\begin{eqnarray}
G^{(1)} &\equiv& - 4 L_{10}^{(0)}(\mu) - {1\over 48 \pi^2} 
\log{M_\pi^2 \over \mu^2} - {1\over 96 \pi^2} 
\log{M_K^2 \over \mu^2} - {1\over 32 \pi^2} \ \ ,
\nonumber \\
G_\pi^{(2)} &\equiv& {M_\pi^2 \over \pi^4 F_\pi^2} \bigg[ 
\log{M_\pi^2 \over \mu^2} 
\left( - {1 \over 288} + {1 \over 768} \log{M_\pi^2 \over \mu^2}
+ {1 \over 768} \log{M_K^2 \over \mu^2} + 
\pi^2 L_{10}^{(0)}(\mu) + {\pi^2 \over 2} L_9^{(0)}(\mu)  \right)
\nonumber \\
& & \phantom{xxxxx} + \log{M_K^2 \over \mu^2} 
\left( - {1 \over 576} - 
{1 \over 1536} \log{M_K^2 \over \mu^2} \right) \bigg] \ \ ,
\label{dmo9} \\
G_K^{(2)} &\equiv& {M_K^2 \over \pi^4 F_\pi^2} 
\log{M_K^2 \over \mu^2} \bigg[ - {17 \over 3072} 
+ {1 \over 1024} \log{M_K^2 \over \mu^2} + 
{\pi^2 \over 2} L_{10}^{(0)}(\mu) + {\pi^2 \over 4} 
L_9^{(0)}(\mu) \bigg] \ \ ,
\nonumber \\ 
G_{\rm NUM}^{(2)} &\equiv& - {49 \over 13824 \pi^4}~ 
{M_\pi^2 \over F_\pi^2} - {5 \over 36864 \pi^4}~ 
{M_K^2 \over F_\pi^2} 
- {1.927 \cdot 10^{-6}~{\rm GeV}^2 \over F_\pi^2} 
+ {4.132 \cdot 10^{-9}~{\rm GeV}^2 \over F_\pi^2} \ \ ,
\nonumber \\
G_{\rm CT}^{(2)} &\equiv& {4 M_\pi^2 \over F_\pi^2} \left( 
Q_{\rm A}^{{\overline{\rm MS}}} (\mu) 
+ R_{\rm A}^{{\overline{\rm MS}}}(\mu) - 
Q_{\rm V}^{{\overline{\rm MS}}} (\mu) 
- R_{\rm V}^{{\overline{\rm MS}}}(\mu) \right) 
+ {8 M_K^2 \over F_\pi^2} \left( 
R_{\rm A}^{{\overline{\rm MS}}} (\mu) 
- R_{\rm V}^{{\overline{\rm MS}}}(\mu) \right) \ \ ,
\nonumber 
\end{eqnarray}
where $G^{(1)}$ and $G^{(2)}$ are respectively the 
ChPT one-loop and two-loop contributions and 
the final two (numerical) terms in $G_{\rm NUM}^{(2)}$ 
arise from the so-called sunset amplitudes.  All the two-loop 
quantities are new contributions which have not previously appeared 
in the literature.  We shall discuss the counterterm contributions 
separately in a later section.

\begin{center}
{\bf Phenomenological Determination of DMO Spectral Integral}
\end{center}

One anticipates an accurate determination of the DMO spectral
integral 
\begin{equation}
I_{\rm DMO} \equiv \int_0^\infty ds ~ 
{(\rho_{{\rm V}3}^{(1)} - \rho_{{\rm A}3}^{(1)})(s) \over s} 
\label{dmo12}
\end{equation}
insofar as the inverse factor of the squared-energy suppresses 
contributions at large $s$. The dominant contributions to 
$I_{\rm DMO}$ will arise from the $2\pi$ and $3\pi$ resonances 
$\rho(770)$ and $a_1(1260)$, with smaller contributions from 
$n_\pi \ge 4$ multipion states, ${\bar K}K$-multipion states, 
{\it etc}.  One can access such information via studies of 
tau-lepton decay and $e^+e^- \to 2\pi, etc$ cross 
sections.~\cite{dg}  

An evaluation of $I_{\rm DMO}$ which employs as input 
the most recent $\tau$ and $e^+e^-$ data available and 
which neglects contributions from $s> m_\tau^2$ yields~\cite{ho} 
\begin{equation}
I_{\rm DMO} = (26.4 \pm 1.5) \cdot 10^{-3} \ \ ,
\label{dmo13}
\end{equation}
with an uncertainty of about $5.7\%$.  In a modified 
analysis, one defines the laplace transformed quantity 
\begin{equation}
{\hat I}_{\rm DMO}(M^2) \equiv \int_0^\infty ds ~ e^{s/M^2}~
{(\rho_{{\rm V}3}^{(1)} - \rho_{{\rm A}3}^{(1)})(s) \over s} 
+ {F_\pi^2 \over M^2} - {C_6 {\cal O}_6 \over 6 M^6} 
- {C_8 {\cal O}_8 \over 24 M^8} \ \ ,
\label{dmo14}
\end{equation}
where $I_{\rm DMO} = \lim_{M \to\infty} {\hat I}_{\rm DMO}(M^2) $ 
and the final two terms involve nonperturbative contributions.  
A fit to several $M^2$-moments then yields~\cite{ho}  
\begin{equation}
I_{\rm DMO} = (25.8 \pm 0.3 \pm 0.1) \cdot 10^{-3} \ \ ,
\label{dmo15}
\end{equation}
with a claimed uncertainty of about $1.24\%$.  

\begin{center}
{\bf Inputs to Counterterm Contributions} 
\end{center}

Consider the contributions in Eq.~(\ref{dmo8}) arising from 
counterterms.  There are those from the chiral 
lagrangians of order $p^4$ ($L_{9}^{(0)}$ and $L_{10}^{(0)}$) 
and those from lagrangians of order $p^6$ 
($Q_{\rm A}^{{\overline{\rm MS}}}$, 
$R_{\rm A}^{{\overline{\rm MS}}}$, 
$Q_{\rm V}^{{\overline{\rm MS}}}$ and 
$R_{\rm V}^{{\overline{\rm MS}}}$).
All these counterterms are evaluated in ${\overline{\rm MS}}$ 
renormalization.\footnote{The reader is referred to Ref.~\cite{gk4} 
for a discussion of renormalization scheme dependence of the 
counterterms.}  For notational consistency with the 
axial propagator counterterms, the vector propagator counterterms, 
defined earlier in Refs.~\cite{gk1,gk2} as $Q$ and $R$, are denoted 
here respectively as $Q_{\rm V}^{{\overline{\rm MS}}}$ and 
$R_{\rm V}^{{\overline{\rm MS}}}$.  

The apparent dependence in Eq.~(\ref{dmo8}) on the scale 
$\mu$ is illusory.  The explicit scale dependence of the logarithmic 
terms is compensated by implicit scale dependence in the $p^4$ and $p^6$ 
counterterms such that the LHS is independent of scale.  In
particular, the scale dependence of the $p^6$ counterterms is 
\begin{eqnarray}
\lefteqn{ \left( Q_{\rm A}^{{\overline{\rm MS}}} (\mu) - 
Q_{\rm V}^{{\overline{\rm MS}}} (\mu) \right) - 
\left( \mu \to \mu_0 \right) =}  
\nonumber \\
& & \phantom{xxxx} {1\over (16 \pi^2)^2} \log{\mu_0^2 \over \mu^2} 
\bigg[ {5\over 32} 
- {3\over 2} \left( L_9^{(0)}(\mu_0) + L_{10}^{(0)}(\mu_0) \right) 
+ {3\over 32} \log{\mu_0^2 \over \mu^2} \bigg]  \ \ ,
\nonumber \\
& & \left( R_{\rm A}^{{\overline{\rm MS}}} (\mu) - 
R_{\rm V}^{{\overline{\rm MS}}} (\mu) \right) - 
\left( \mu \to \mu_0 \right) = 
\label{dmo9a} \\
& & \phantom{xxxx} {1\over (16 \pi^2)^2} \log{\mu_0^2 \over \mu^2} 
\bigg[ {17\over 96} - {1\over 2} \left( L_9^{(0)}(\mu_0) 
+ L_{10}^{(0)}(\mu_0) \right) + {1\over 32} 
\log{\mu_0^2 \over \mu^2} \bigg]  \ \ ,
\nonumber 
\end{eqnarray}
as can be obtained from renormalization group equations~\cite{gk1}.  

The collection of $p^4$ and 
$p^6$ counterterms originate from the renormalization procedure, 
and each must ultimately be determined from experiment.  
The $p^4$ counterterms $L_{9}^{(0)}$ and $L_{10}^{(0)}$ are 
already known~\cite{eck94} from {\it one-loop} analyses to accuracies 
respectively of about $10\%$ and $13\%$,
\begin{eqnarray}
L_{9}^{(0)}(M_\rho) &=& \ \ (6.9 \pm 0.7) \cdot 10^{-3} \qquad 
({\rm from} \ \langle r^2 \rangle^\pi_{\rm V}) \ \ ,
\nonumber \\
L_{10}^{(0)}(M_\rho) &=& -(5.5 \pm 0.7) \cdot 10^{-3}  \qquad 
({\rm from}\ \pi^- \to e {\bar \nu}_e \gamma) \ \ .
\label{dmo10} 
\end{eqnarray}
We may employ these values for $L_{9}^{(0)}$ and $L_{10}^{(0)}$ 
in the two-loop quantities $G_\pi^{(2)}$ and $G_K^{(2)}$ 
of Eq.~(\ref{dmo8}) because any error made is of still higher order.  
This is, however, not true for the $L_{10}^{(0)}$ dependence in 
$G^{(1)}$, for which consistency demands a two-loop evaluation.  
That is, suppose hypothetically that a physical amplitude 
existed which contained $L_{10}^{(0)}$, but no other counterterms 
to either one-loop or two-loop order.  A phenomenological determination of 
$L_{10}^{(0)}$ carried out in a two-loop analysis would yield 
a value somewhat different from that obtained in a one-loop 
analysis.  We would expect this shift in $L_{10}^{(0)}$ 
to be of the same order as the $p^6$ counterterms which comprise 
$G_{\rm CT}^{(2)}$ in Eq.~(\ref{dmo9}), and so we are not 
justified {\it a priori} in ignoring it.  

Of the $p^6$ counterterms, we have previously used spectral 
function sum rules to obtain the estimates, 
\begin{eqnarray}
Q_{\rm V}^{{\overline{\rm MS}}}(M_\rho) &=& (3.7 \pm 2.0) \cdot 10^{-5}
\qquad ({\rm Ref.}~\cite{gk2})  \ \ ,\nonumber \\
Q_{\rm A}^{{\overline{\rm MS}}}(M_\rho) &=& (1.4 \pm 0.5) \cdot 10^{-4}
\qquad ({\rm Ref.}~\cite{gk4})  \ \ ,
\label{dmo11} 
\end{eqnarray}
where the result of Ref.~\cite{gk4} has been scaled down to 
$\mu = M_\rho$ in the above.  The relatively large uncertainties 
in each of these reflects the paucity of existing data in the 
hypercharge channel.  It is also possible to obtain estimates of
certain counterterms via the assumption of resonance 
exchange saturation~\cite{etc}.  Thus, a reanalysis~\cite{kk} of the paper 
by Jetter~\cite{je} on $\gamma \gamma \to \pi \pi$ and $\eta \to 
\pi \gamma\gamma$ yields 
\begin{equation} 
Q_{\rm V}^{{\overline{\rm MS}}}|_{\rm res} 
= F_\pi^2 {C_S^\gamma C_S^m \over M_S^2}
\ \simeq \  \pm 3.0 \cdot 10^{-5} \ ,
\qquad \qquad R_{\rm V}^{{\overline{\rm MS}}}|_{\rm res}  = 0 \ \ .
\label{dmo16}
\end{equation}
Since the counterterm $Q_{\rm V}^{{\overline{\rm MS}}}$ receives
contributions in the resonance exchange approach only from scalar 
exchange, its magnitude (the sign is not fixed) is rendered 
small.  The resonance couplings $C_S^\gamma$ and $C_S^m$ are defined 
in Appendix D.2 of Ref.~\cite{BGS94}, and the result is seen to agree 
nicely with our sum rule determination cited in Eq.~(\ref{dmo11}).  
As for $R_{\rm V}^{{\overline{\rm MS}}}$, 
although the contributions from low-lying resonances vanish, there 
will of course be continuum contributions.  Even so, we disagree 
with the numerical estimates given in Ref.~\cite{je}, and the 
vanishing result for $R_{\rm V}^{{\overline{\rm MS}}}|_{\rm res}$ 
implies that the interpretation of a large $\eta \to \pi \gamma\gamma$
amplitude given in Ref.~\cite{je} is not tenable.   

\begin{center}
{\bf Analysis and Conclusions}
\end{center}

We are now ready to study the numerical implications of the DMO sum rule.
To begin, let us temporarily avoid any use of counterterm inputs 
in order to get a result free of theoretical errors.  Then 
using the conservative estimate of Eq.~(\ref{dmo13}) for the DMO 
spectral integral $I_{\rm DMO}$, we obtain from the DMO sum rule 
the numerical constraint
\begin{eqnarray}
& & L_{10}^{(0)}(M_\rho)
-{M_\pi^2 \over F_\pi^2} 
\left[ Q_{\rm A}^{{\overline{\rm MS}}}(M_\rho)-
       Q_{\rm V}^{{\overline{\rm MS}}}(M_\rho)\right]
-{M_\pi^2+2 M_K^2 \over F_\pi^2} 
\left[ R_{\rm A}^{{\overline{\rm MS}}}(M_\rho)-
       R_{\rm V}^{{\overline{\rm MS}}}(M_\rho)\right]
\nonumber \\
& & = -{1\over 4}\left[ I_{\rm DMO} - {\overline G}^{(1)}(M_\rho) 
- {\overline G}^{(2)}(M_\rho) \right] = 
-0.0038 \pm 0.0004 \ \ ,
\label{dmo17}
\end{eqnarray}
where ${\overline G}^{(1)} \equiv G^{(1)} + 4 L_{10}^{(0)}$ and 
${\overline G}^{(2)} \equiv G^{(2)} - G_{\rm CT}^{(2)}$.  
The RHS of Eq.~(\ref{dmo17}) will change in a known manner 
as the scale $\mu$ changes but the error bar, due exclusively to 
the uncertainty in the evaluation of $I_{\rm DMO}$, will remain 
the same.  The value of the RHS is seen to be rather smaller in magnitude 
than the one-loop value $L_{10}^{(0)}(M_\rho) \simeq -0.0055$.  
This is due in large part to sizeable chiral logarithms present 
in $G_\pi^{(2)}$ and $G_K^{(2)}$.  That is, 
the numerical contribution of $I_{\rm DMO}$ to the 
RHS (equal to $-0.0066$) is reduced via 
cancellation with the one-loop term ${\overline G}^{(1)}$ 
to $-0.0054$ and then via cancellation with the two-loop term 
${\overline G}^{(2)}$ to the final value of $-0.0038$.  The 
latter reduction is about a $30\%$ effect. 

Can the contribution of $L_{10}^{(0)}(M_\rho)$ be disentangled 
from the mass corrections on the LHS of Eq.~(\ref{dmo17})?  
The interest in this question is underlined by the large 
numerical effect, just mentioned, the chiral logarithms have on the LHS of 
the DMO sum rule. From Eq.~(\ref{dmo17}), one can already conclude that if 
$L_{10}^{(0)}(M_\rho)$ stays close to the one-loop determination 
cited in Eq.~(\ref{dmo10}), the mass corrections involving the 
$p^6$ counterterms have to be both substantial and negative at scale
$\mu=M_\rho$. Conversely, if these mass corrections turn out to be 
small at $\mu = M_\rho$, any two-loop analysis would need to 
yield a determination for $L_{10}^{(0)}(M_\rho)$ reduced by about 
$30\%$ relative to its one-loop value.  

The information gathered in Eqs.~(\ref{dmo11}),(\ref{dmo16}) is not 
by itself sufficient to conclusively answer this question. 
Nevertheless it is instructive 
to investigate in more detail the individual contributions to the LHS
of the DMO sum rule. Using the estimates of Eq.~(\ref{dmo11}), the 
contribution of the Q-type counterterms is numerically found to be
\begin{equation}
-{4 M_\pi^2 \over F_\pi^2}
\left[ Q_{\rm A}^{{\overline{\rm MS}}}(M_\rho)-
       Q_{\rm V}^{{\overline{\rm MS}}}(M_\rho)\right] 
= (9.3 \pm 5.0) \cdot 10^{-4}
\label{Qnum}
\end{equation}
where we have added errors on $Q_{\rm A}^{{\overline{\rm MS}}}$ and
$Q_{\rm V}^{{\overline{\rm MS}}}$ in quadrature. The Q-terms thus
contribute approximately $3.5\%$ to the value obtained 
phenomenologically for the RHS of the DMO sum rule. 
For $R_{\rm V}^{{\overline{\rm MS}}}$ we can employ the resonance saturation
estimate of Eq.~(\ref{dmo16}). No corresponding estimate is 
available, however, for $R_{\rm A}^{{\overline{\rm MS}}}$. 
{\it A priori}, such contributions proceed through exchange of 
scalar resonances as well, but the couplings 
of those resonances to two axialvector currents are not known experimentally. 
However, since both R-type counterterms are Zweig-rule suppressed, 
we expect these to be small compared to the Q-type counterterms. 

There is, however, a well-known drawback of such resonance saturation
estimates of counterterms. The method does not specify the scale at 
which the counterterm has to be taken. The generally accepted
procedure is to employ the resonance mass as the relevant scale and to use 
the variation of the counterterm in a range between, say, $\mu= 0.5
\to 1~{\rm GeV}$ as an estimate of the uncertainty of the method. In the DMO
sum rule, the counterterms turn out to vary strongly with scale. This can
be seen either by using Eqs.~(\ref{dmo9a}) or be read off from Table 1
where we give the two-loop contributions $G_\pi^{(2)}$, $G_K^{(2)}$
and $G_{\rm NUM}^{(2)}$ for three different scales $\mu$. 

\begin{center}
\begin{tabular}{l|ccc}
\multicolumn{4}{c}{Table~1: {Scale dependence}} \\
\hline\hline
 $\mu$~(GeV)   & $0.50$ & $0.77$ & $1.00$ \\ \hline
$G_\pi^{(2)}(\mu)$       & $\ \ 0.0018$  & $\ \ 0.0023$  & $\ \ 0.0026$   \\
$G_K^{(2)}(\mu)$         & $\ \ 0.0001$  & $\ \ 0.0042$  & $\ \ 0.0065$   \\
$G_{\rm NUM}^{(2)}$      & $-0.0003$ & $-0.0003$ & $-0.0003$  
\\ \hline 
Sum$(\mu)$               & $\ \ 0.0016$  & $\ \ 0.0062$  & $\ \ 0.0088$   \\ \hline
\end{tabular}
\end{center}

For the range of scales between $\mu=0.5~{\rm GeV}$ and $\mu=1~{\rm GeV}$ the
variation is $0.007$, or $27\%$ of the RHS of the DMO sum rule.
This variation with scale is of course counterbalanced by the 
$p^6$ counterterms. However, as just explained, if an estimate for 
these counterterms is used where one has no control over the scale 
at which the counterterm is fixed, the strong
variation with scale translates into a large uncertainty in the
determination of the LHS of the DMO sum rule. 
Consequently, it will be possible to disentangle $L_{10}^{(0)}(M_\rho)$
from the mass corrections on the LHS of the DMO sum rule only if one 
can perform an independent determination of 
$R_{\rm A}^{{\overline{\rm MS}}}-R_{\rm V}^{{\overline{\rm MS}}}$
directly from physical observables. 

To close the discussion, we present some remarks concerning 
existing and future work in this direction. A related process in which 
$L_{10}^{(0)}$ contributes at order $p^4$ is the radiative pion decay 
$\pi\rightarrow e {\bar \nu}_e \gamma$. In this process, however, it is the 
combination $L_9^{(0)}+L_{10}^{(0)}$ which occurs, and the structure dependent
part of the amplitude is actually used to determine this combination 
of $p^4$ counterterms.\footnote{In SU(2) notation, the relevant 
combination of counterterms is $2 l_5^r-l_6^r$.  Its relation to 
the SU(3) couplings is known only to one-loop order.~\cite{gl2} 
A direct comparison with our analysis is therefore not yet possible.}
At order $p^6$, additional counterterms will 
contribute, some of them not occuring in the two-point functions 
considered here. Bijnens and Talavera have calculated the radiative 
pion decay amplitude to two-loop order~\cite{BT97}, although in chiral 
SU(2).  Using the hypotheses of resonance saturation 
(with just vector- and axialvector resonances), they estimate the
$p^6$ counterterm contributions to be small. The sum of the 
two-loop contributions is large, however, leading to an enhancement of
about $15\%$ for $2 l_5^r-l_6^r$ at the scale $M_\rho$. The advantage of 
an SU(2) calculation is that mass corrections are 
always suppressed by factors of $M_\pi^2$. The large uncertainties 
due to scale-dependence of resonance saturated counterterms found 
in our SU(3) calculation are thus largely avoided.  The comparsion of
our results to the DMO sum rule in its SU(2) version will thus be an
interesting avenue for future work. On the other hand, working in
chiral SU(3) offers the possibility of studying additional processes with
external currents/Goldstone Bosons containing strangeness. For
instance, there are the radiative $K_{l2}$ decays~\cite{BEG93}, where more 
precise data is expected in the near future~\cite{Daphne}.  However, 
because at one-loop order $K \to \ell {\bar \nu}_\ell \gamma$ involves the 
combination $L_9^{(0)} + L_{10}^{(0)}$, learning more
about just $L_{10}^{(0)}$ will require the study of other processes 
in which $L_9^{(0)}$ enters separately. The meson form factors 
of vector current matrix elements, $\langle P|V_\mu|Q\rangle$, 
where $P, Q = \pi,K,\eta$, offer one possibility to 
study mass corrections to $L_9^{(0)}$.  Partial results on a combination of 
such form factors have been reported recently~\cite{PS97}.   It 
remains to be seen whether the mass corrections to the individual 
form factors can be disentangled from the contributions of 
the low energy constant $L_9^{(0)}$. 

\vspace{1.6cm}

To conclude, we have used our recent two-loop calculation of isospin and 
hypercharge vector and axialvector current propagators in chiral 
perturbation theory to re-analyze the DMO sum rule.  The resulting 
relation, valid to second order in the light quark masses, 
is summarized in Eqs.~(\ref{dmo7})-(\ref{dmo9}).   Since we 
work in SU(3)$\times$SU(3) chiral symmetry, both pion and 
kaon mass dependence is present.  A phenomenological determination 
of the DMO spectral integral yielded a numerical constraint on the 
combination of $p^4$ and $p^6$ counterterms, given in
Eq.~(\ref{dmo17}), and prospects for extending the analysis 
were considered.  As regards the overall question of how 
quickly the chiral expansion is converging, we find 
individual contributions at two-loop level to occur at roughly the 
$30\%$ level.

\acknowledgments 
The research described here 
was supported in part by the National Science Foundation 
and by Schweizerischer Nationalfonds.  We acknowledge 
useful conversations with J. Bijnens and J. Donoghue.  
We thank M. Knecht for allowing us to use an unpublished 
result~\cite{kk}.


\eject

\end{document}